\documentclass[10pt,twocolumn]{article}
\usepackage{a4wide}
\usepackage{epsfig}
\usepackage{booktabs}
\usepackage{amssymb}
\usepackage[T1]{fontenc}
\newfont{\xLfont}{pzcmi8r at 10pt}
\newcommand{\xLoops}{{\xLfont xloops}}
\newcommand\rbksection[1]{\section{\sffamily\bfseries #1}}

\date{December 7, 1998}
\author{Richard Kreckel\footnote{Inst. f. Physik, 
Johannes Gutenberg-Univ. Mainz, Germany, email: {\tt richard.kreckel@uni-mainz.de}}}
\title{
\begin{flushright}\null\vskip-35mm
{\normalfont\normalsize MZ-TH/98-54}
\vskip 10mm
\end{flushright}
{\sffamily\bfseries Parallelization of adaptive MC integrators \\ 
Recent pvegas developments}}

\begin{document}

\maketitle

\begin{abstract}
  \small This paper shortly describes some important changes to the
  \verb|pvegas|-code since its first publication. It proceeds with a
  report on the scaling-behavior that was found on a wide range of
  current parallel hardware and discusses some issues of optimization
  that may be thrown up.
\end{abstract}

\vskip 5mm Since the first public announcement of our parallel version
of G.~P.~Lepage's \verb|vegas|-algorithm~\cite{Lepage78} for
multidimensional numerical integration in October
1997~\cite{pvegasPaper} work has been done to improve the
\verb|pvegas|-code\footnote{The sources are available via
  anonymous-FTP from {\tt
    ftp://ftpthep.physik.uni-mainz.de/pub/pvegas/}.} by adding new
features and making it more portable such that nearly all present-day
parallel hardware is supported.\footnote{Independently from us,
  Sini\v{s}a Veseli has worked on an implementation of {\tt vegas} for
  machines with distributed memory using the
  PVM-library~\cite{VegasVeseli} and Thorsten Ohl has prepared a
  parallel version for MPI-based systems that makes an attempt to
  overcome {\tt vegas}' inherent problems with non-factorizable
  singularities~\cite{VegasOhl}.} We feel that these improvements
ought to be published to the community of \verb|pvegas|-users now
especially since we finally consider this work completed due both to
positive reports by users and a convergence to the widely accepted
standards Posix 1003.1c and MPI~\cite{MPI}.

Our original approach to parallelization based on splitting up the
\(D\)-dimensional space into a \(D_\parallel\)-dimensional parallel
subspace and its orthogonal complement (the
\(D_\perp\!=\!D\!-\!D_\parallel\) -dimensional orthogonal space) and using the
stratification-grid for decomposition as outlined
in~\cite{pvegasPaper} remains untouched.
Section~\ref{sec:improvements} describes the most important additions.

Simultaneously \verb|pvegas| has been applied by numerous researchers
both in industry and universities. Its scaling-behavior has been
probed in practice on a wide spectrum of hardware. The remaining two
sections are devoted to a discussion of platforms.

\rbksection{Code improvements}\label{sec:improvements}
Since the original \verb|pvegas| features an independent random number
generator (RNG) for each processor in order to reduce the sequential
fraction in Amdahl's law, the numerical output is subject to
additional random fluctuations.  Although these fluctuations are
nothing new (merely representing the statistical nature of MC
integration) they can nevertheless trouble. It turned out to be
impossible to obtain exactly the same output if more than one
processor was used.  To get around this problem the code was
restructured in order to allow for reproducible results if the user
wishes to.  If enabled, this feature will initialize all RNGs
identically and subsequently an algorithm will decide how much to
advance each RNG such that exactly the same \(D\)-dimensional
sample-points are evaluated. We will call this feature causal
random-number generation. Thus the output is independent from the
number of processors \(p\) and can easily be checked by a machine,
using \verb|diff| for instance. It is self-evident that this slows
down the program and should be used for the purpose of debugging only
and not in production runs.\footnote{Since the code for this feature is
  completely hash-defined, absolutely no overhead is introduced when it
  is switched off, thus guaranteeing the usual performance.}

The second innovation is an implementation of \verb|pvegas| using the
MPI Message-Passing Interface Standard~\cite{MPI}. This enlarges the
spectrum of machines capable of running \verb|pvegas| substantially,
including massive parallel machines and networks of workstations
(NOWs) alike. This MPI-code also allows causal random number generation.

\begin{table*}[t]
\label{tab:machinery}
%\begin{tabular*}{\textwidth}{@{}l@{\extracolsep{\fill}}l@{\extracolsep{\fill}}l@{\extracolsep{\fill}}l@{\extracolsep{\fill}}l@{\extracolsep{\fill}}r@{\extracolsep{\fill}}l@{\extracolsep{\fill}}l@{}}
\begin{tabular*}{\textwidth}{@{}l@{\extracolsep{\fill}}l@{\extracolsep{\fill}}l@{\extracolsep{\fill}}l@{\extracolsep{\fill}}l@{\extracolsep{\fill}}r@{\extracolsep{\fill}}l@{\extracolsep{\fill}}l@{}}
\toprule
Vendor: & Architecture: & CPU: & MHz: & OS: & \(p_{max}\): & Model: & comment: \\[1mm]
\midrule
Convex & SPP-1200 & PA-7200 & 120 & SPP-UX 4.2 & 46 & CPS & suffers from a looping \\[-1mm]
 & & & & & & & main thread of execution \\[2mm]
HP & X-class & PA-8000 & 180 & SPP-UX 5.2 & 46 & CPS/ & dto., but only if \\[-1mm]
 & & & & & & Posix & CPS-threads are used \\[2mm]
Cray & T3D & EV4 & 150 & {\sc Unicos Max} & 256 & MPI & dto., since MPI (with\\[-1mm]
 & & & & 1.3.0.3 & & & explicit master-process) \\[2mm]
Siemens- & Solaris-NOW & Pentium-II & 300 & SunOS 5.6 & 31 & MPI & dto. (prototype of \\[-1mm]
Scali-Dolphin & & & & & & & commercial product) \\[2mm]
DEC & AlphaServer 8400 & EV5 & 300 & D.U. 4.0 & 6 & Posix & \\[-1mm]
 & (``Turbo-Laser'') & & & & & & \\[2mm]
SGI & Origin 200 & R10000 & 180 & IRIX 6.4 & 4 & Posix & \\[2mm]
Sun & E3000 & UltraSparc & 250 & SunOS 5.5.1 & 4 & Posix & \\[2mm]
self-made & Linux-NOW & AMD K6 & 233 & GNU/ & 5 & MPI & plus one additional \\[-1mm]
system & & & & Linux 2.0 & & & dedicated master-machine \\
\bottomrule
\end{tabular*}
\caption{Overview of Hardware tested. \(p_{max}\) refers to the 
number of CPUs which were used in our tests, not necessarily the 
number of CPUs installed.}
\end{table*}

\rbksection{New platforms since 1996}\label{sec:newplatforms}
Several new hardware-platforms have become available since our first
tests of \verb|pvegas| in Winter 1996/97~\cite{pvegasPaper}. While the
very high-end of supercomputers remains a domain of massively parallel
machines~\cite{Top500}, SMP-workstations with two or more CPUs are
quickly finding their way into labs or even on the desktops of
individual researchers. Since these machines feature the
programming-paradigm of shared-memory and in most cases support
Posix-threads as defined by IEEE in {\sc Posix} Section 1003.1c, they
are well-suited for running the multi-threaded \verb|pvegas| and we
report on some of them in the present paper. In addition, the newer
MPI version using the same decomposition method as the approved
\verb|pvegas| has been tested on Cray-systems and NOWs.  These results
will also be reviewed in the following section.

We challenge the selection of machines in Table~\ref{tab:machinery}
with the computational problem already familiar from our original
publication~\cite{pvegasPaper}. (For this article we include only
machines that have at least 4 CPUs.) The list includes commercially
available products (the SMP machines AlphaServer 8400 and E3000, the
mixed architectures SPP-1200, X-class and Origin 200 and massive
parallel systems like the T3D) as well as systems built from
commodity hardware like the Paderborn cluster of 32 Dual Pentium-II
with a Scalable Coherent Interconnect (SCI) and a quickly assembled
Linux-NOW running MPI over ordinary Ethernet.

\begin{figure*}[t]
  \hspace{0.053\textwidth}
  \epsfig{file=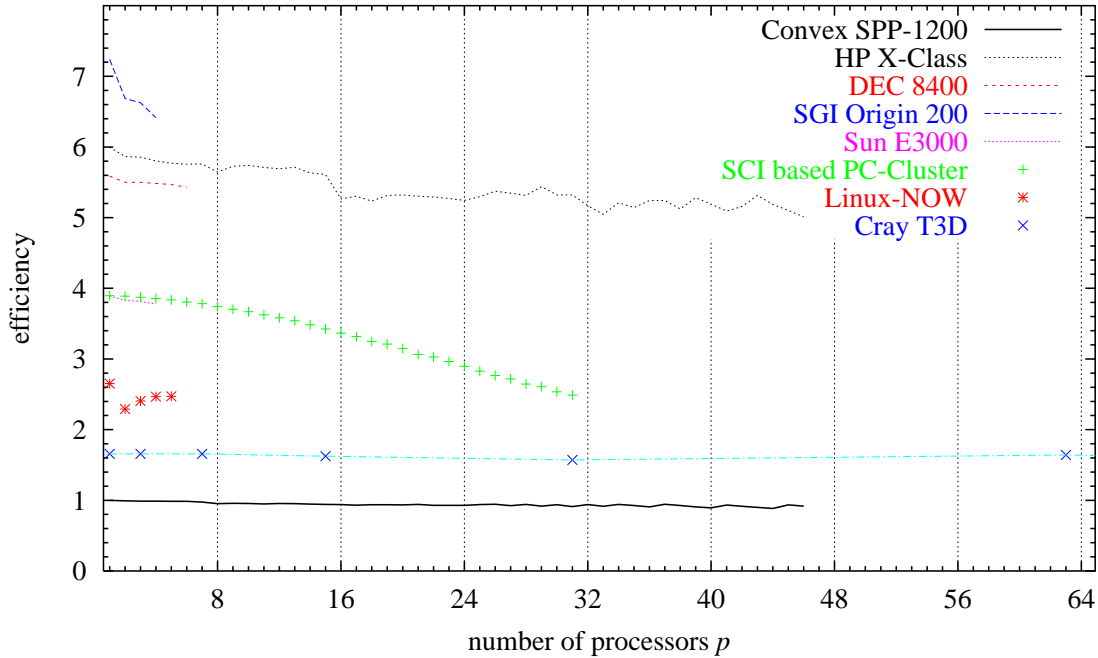,angle=270,width=0.86\textwidth}
  \caption{\label{fig:scalings}Relative efficiency of {\ttfamily pvegas} on different architectures, normalized to the SPP-1200.} 
\end{figure*}

\begin{figure*}[t]
  \hspace{0.053\textwidth}
  \epsfig{file=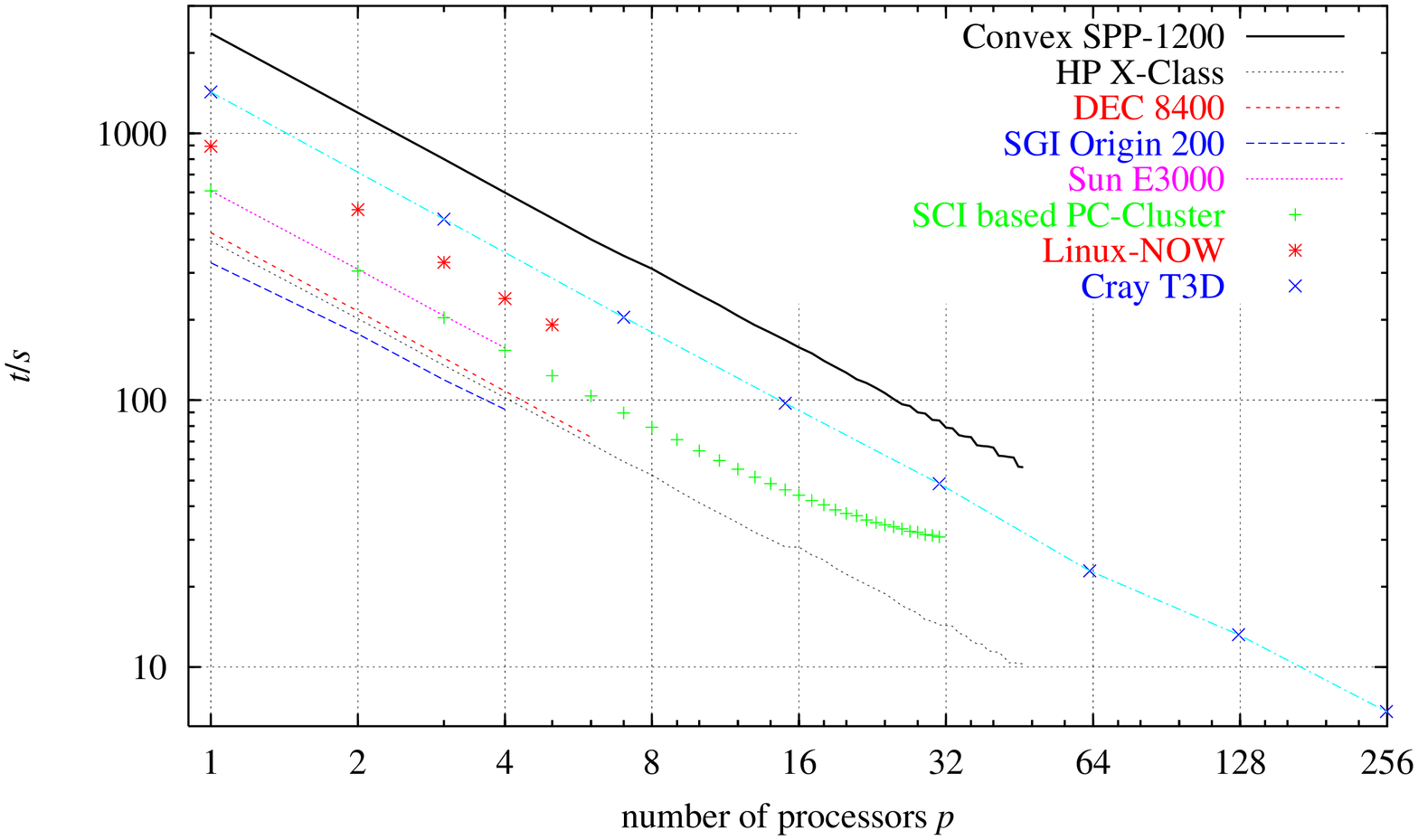,angle=270,width=0.86\textwidth}
  \caption{\label{fig:runtimes}Runtimes of {\ttfamily pvegas} on different architectures in seconds.} 
\end{figure*}

% Test with 19 CPUs on SCI:
% 3.5134 as many points took only 120.19s, 
% since 120.19/38.84 = 3.0945 the drop-off is mostly gone!

\rbksection{Comparison}
To recapitulate, the challenge consisted of integrating a normalized
test function which demanded evaluation of 8 Dilogarithms computed
with a method outlined in~\cite{TVDilogPaper}. (This case resembles a
typical situation in \xLoops~\cite{xLoopsIntro}.)  One must,
however, be careful when deducing anything for other
integrals---before embarking on large-scale computations one should
always consider measuring the behavior of one's machine.

The \(D\!=\!5\) -dimensional problem is split up into a
\(D_\parallel\!=\!2\) -dimensional parallel space and a
\(D_\perp\!=\!3\) -dimensional orthogonal space. Trying to integrate
\(8.2\cdot10^6\) sample-points results in a grid with 21 slices in
each dimension. Thus \(2\cdot21^5 = 8\,168\,202\) points are being
evaluated.

This integration idealizes a realistic calculation in elementary
particle physics and it turned out to be quite able to probe the
hardware structure and uncover problems in configuration and
optimization. This is the reason why we continued using it for
measurements of efficiency.

In Fig.~\ref{fig:scalings} we choose to normalize all measured
efficiencies with respect to one processor on the Convex SPP-1200
which took about 40 minutes to complete the task.
Fig.~\ref{fig:runtimes} shows absolute runtimes on the hardware tested
plotted double-logarithmically.

Both figures demonstrate the rather good overall scalability. Several
aspects deserve special mention:
\begin{itemize}
\item All curves are modulated by a visible grain-size effect at
  \(p\gtrsim32\) as can best be seen at the X-class. This is to be
  expected by the nature of the test.
\item A slight drop-off at the boundary of hypernodes is apparent in
  the case of the X-class (multiples of 16) and can also be found in
  the runtimes of the SPP-1200 (multiples of 8).
\item The SCI-based cluster of PCs' saturation is due to problems with
  sub-optimal choices of SCI parameters in this prototype machine---a
  production-machine can be expected to perform better. Tests have
  shown that on this machine the saturation can in principle be
  avoided by using larger problem-sizes.
\item The T3D's drop-off can almost purely be explained by the
  grain-size effect. Even a decrease of grain-size by increasing
  \(D_\parallel\) from 2 to 3 still showed nearly perfect scaling up
  to \(p\!=\!256\).
\end{itemize}

The good scalar performance of the SGI Origin 200 and the X-class is
somewhat surprising. It resembles most benchmarks (e.g.~\cite{Top500},
based on {\sc Linpack}~\cite{Linpack}) only partially.  Of course
these differences can directly be traced back to the numerical effort
of our integrand. Inspection of the code with the profiling-tool
available on each individual system reveals some sources for the
strikingly different performances---the ability of the processor to
deal with code full of jumps together with the compiler's performance
at optimization.

To understand this we need to have a look at the benchmark integrand.
The series for calculating Dilogarithms suggested
in~\cite{TVDilogPaper} and used here is enhanced by applying some
relations holding between Dilogarithms in order to evaluate the series
only where it converges (within the complex unit-circle) and further
where it converges fast (within a rectangle inside the unit-circle
that not even touches the circle itself). This transformation with
many conditional statements and even occasional recursions is in
itself the first source of possible performance losses. It clearly
probes the processor's branch-prediction and branch-penalty together
with its potential in doing out-of-order execution---the latter is
probably the cause for the relatively poor scalar performance of the
Ultra-SPARC processors.  The result of the outlined transformation, a
sum of Taylor series, involves complex multiplications, a code
optimized much better by the IRIX compiler than on all the other ones
(e.g.  effectively four cycles per complex multiplication in contrast
to 27 on the Turbo-Laser).\footnote{Only the vendor's compilers with
  aggressive optimization-settings were used on these platforms:
\begin{itemize}
\item X-class: Exemplar {\tt c89} C-compiler V 1.2.1\\[-4ex]
\item Origin 200: MIPSpro C-compiler V 7.20\\[-4ex]
\item E3000: Sun WorkShop C-compiler V 4.2\\[-4ex]
\item Turbo-Laser: DEC C V5.2-038
\end{itemize}
GNU {\tt gcc} V 2.8.1 was used on the PC-cluster since it turned out
to be the fastest compiler for that particlular code.} The lesson from
all this is that one must be extremely careful when trying to predict
the performance of any nonlinear code.  With respect to \verb|pvegas|
we should of course not use the presented absolute numbers to rate the
genuine \verb|pvegas|-performance.

\rbksection{Conclusion}
The nearly constant scaling of the multi-threaded \verb|pvegas| found
in our original work~\cite{pvegasPaper} on the SPP-1200 turns out to
be a case reached with most current hardware. This was reported
independendly by a wide variety of researchers. While the old SPP-1200
disappoints by a lack of Posix-threads (resulting in the need of a
somewhat clumsy code that respects two different thread-models) the
X-Class provides both of them. Some small and less costly SMP-machines
also do a reasonable job, thus becoming a very attractive tool for the
numerically demanding researcher.

In the meantime, using an explicit farmer-worker model and the
MPI standard for message-passing was found to deliver very satisfying
performance if the price of an idling master CPU can be paid. These
results depend, of course, strongly on the latencies of the
underlying network or message-passing hardware.

\vspace{4ex}
\noindent{\sffamily\bfseries Acknowledgements:} I wish to thank the 
Edinburgh Parallel Computing Centre for hospitality during a visit in
winter 1998 and their invaluable support of part of this work. I am
also grateful to GIP\ Mainz and to Paderborn Center for parallel
Computing for their feedback as well as access to their machines.
Again, Markus Tacke has contributed some very fruitful comments about
issues of optimization.

\end{document}